\documentclass[11pt]{article}
\usepackage{times}
\usepackage{epsfig}
\usepackage{fullpage}
\usepackage{amssymb}

\newcommand{\be}{\begin{equation}}
\newcommand{\ee}{\end{equation}}
\newcommand{\bea}{\begin{eqnarray}}
\newcommand{\eea}{\end{eqnarray}}

\newtheorem{conjecture}{Conjecture}

\newcommand{\qed}{\hfill \rule{1ex}{1ex}\medskip\\} 
\newtheorem{theorem}{Theorem} 
\newtheorem{lemma}[theorem]{Lemma} 
\newtheorem{proposition}[theorem]{Proposition} 
 
\newenvironment{proof}{\paragraph{Proof}}{\qed}

\def\S{Section~}
\let\hat\widehat

\def\cI{{\cal I}}
\def\cE{{\cal E}}

\newcommand\E{\textsf{E}}

\let\Var\var

\newcommand\Prp[1]{\Pr\!\left[{{#1}}\right]}
\newcommand\Ep[1]{\E\left[{{#1}}\right]}
\newcommand\Varp[1]{\Var\left[{{#1}}\right]}

\newcommand\req[1]{(\ref{#1})}
\newcommand\Sref[1]{\S\ref{#1}}
\newcommand\Fref[1]{Figure~\ref{#1}}
\newcommand\tref[1]{~\ref{#1}}
\newcommand\eps\varepsilon

\title{Priority sampling estimating arbitrary subset sums}

\author{Nick Duffield\thanks{All authors are
researchers at AT\&T Labs---Research, Shannon Laboratory,
180 Park Avenue, 
NJ 07932, USA (email: \texttt{(duffield,lund,mthorup)@research.att.com}).}
\hspace{2cm} Carsten Lund$^*$\hspace{2cm} Mikkel 
Thorup$^*$\\[2ex]AT\&T Labs---Research}

\begin{document}
\maketitle 

\begin{abstract}
  Starting with a set of weighted items, we want to create a generic sample of
  a certain size that we can later use to estimate the total weight of
  arbitrary subsets. Applied to internet traffic analysis, the items
  could be records summarizing the flows of packets streaming by a router, with,
  say, a hundred records to be sampled each hour. A subset could be flow
  records of a worm attack whose signature is only determined after
  sampling has taken place. The samples taken in the past 
  allow us to trace the history of the attack even though the worm was
  unknown at the time of sampling.
  
  Estimation from the samples must be accurate even with heavy-tailed
  distributions where most of the weight is concentrated on a few
  heavy items. We want the sample to be weight sensitive, giving
  priority to heavy items. At the same time, we want sampling without
  replacement in order to avoid selecting heavy items multiple times.
  To fulfill these requirements we introduce priority sampling, which
  is the first weight sensitive sampling scheme without replacement
  that is suitable for estimating subset sums.  Testing priority
  sampling on Internet traffic analysis, we found it to perform orders
  of magnitude better than previous schemes.

  Priority sampling is simple to define and implement: we
  consider a steam of items $i=0,...,n-1$ with weights $w_i$. For each
  item $i$, we generate a random number $\alpha_i\in (0,1)$ and create
  a priority $q_i=w_i/\alpha_i$.  The sample $S$ consists of the $k$
  highest priority items. Let $\tau$ be the $(k+1)^{th}$ highest
  priority. Each sampled item $i$ in $S$ gets a weight estimate $\hat
  w_i=\max\{w_i,\tau\}$, while non-sampled items get weight estimate
  $\hat w_i=0$.

  Magically, it turns out that the weight estimates are unbiased, that
  is, $\E[\hat w_i]=w_i$, and by linearity of expectation, we get unbiased
  estimators over any subset sum simply by adding the sampled weight
  estimates from the subset. Also, we can estimate the variance of the
  estimates, and find, surprisingly, that the covariance between estimates
  $\hat w_i$ and $\hat w_j$ of different weights is zero.

  Finally, we conjecture an extremely strong near-optimality; namely that for
  any weight sequence, there exists no specialized scheme for
  sampling $k$ items with unbiased weight estimators that gets smaller
  total variance than priority sampling with $k+1$ items.
  Very recently, Szegedy has settled this conjecture.
\end{abstract}


\section{Introduction}
Starting with a set of weighted items, we want to create a generic
sample of a certain size that we can later use to estimate the total
weight of arbitrary subsets.  Applied to internet traffic analysis,
the items could be records summarizing the flows streaming by a
router, with, say, a hundred records sampled each hour.  A subset
could be flow records of a worm attack whose signature is only
determined after sampling has taken place. The samples taken in the
past allow us to trace the history of the attack even though the worm
was unknown at the time of sampling.

Estimation from the samples must be accurate even with heavy-tailed
distributions where most of the weight is concentrated on a few
heavy items. We want the sample to be weight sensitive, giving
priority to heavy items. At the same time, we want sampling without
replacement in order to avoid selecting heavy items multiple times.
To fulfill these requirements we introduce priority sampling, which
is the first weight sensitive sampling scheme without replacement
that is suitable for estimating subset sums.  Testing priority
sampling on Internet traffic analysis, we found it to perform orders
of magnitude better than previous schemes.

\subsection{Priority Sampling}
\begin{figure}[t]
\begin{center}
\leavevmode
\epsfig{file=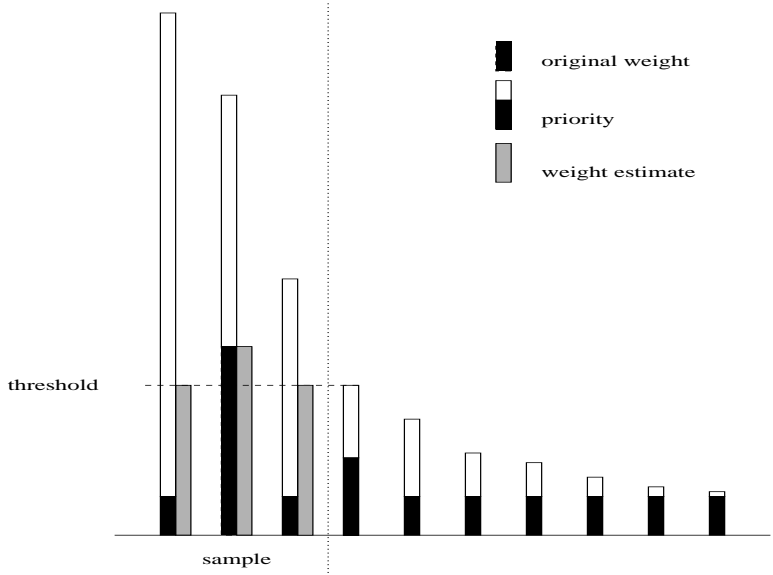,width=4in,height=3in}
\end{center}
\caption{Priority sampling of size 3 from a set of 10 
weighted items.}\label{fig:sample}
\end{figure}

{\em Priority sampling\/} is a fundamental new technique to
sample $k$ items from a stream of weighted items so as to later
estimate arbitrary subset sums. The scheme is illustrated in
Figure~\ref{fig:sample}.
We consider a stream of items with positive weights
$w_0,...,w_{n-1}$. For each item $i=0,..,n-1$, we generate an
independent uniformly random $\alpha_i\in(0,1)$, and a {\em
priority\/} $q_i=w_i/\alpha_i$.   Assuming that all priorities are distinct, the {\em priority
sample $S$ of size $k<n$\/} consists of the $k$ items of highest
priority. An associated {\em threshold\/} $\tau$ is the
$(k+1)^{\mathrm{th}}$ priority.  Then $i\in S\iff q_i>\tau$. Each
sampled item $i\in S$ gets a weight estimate $\hat
w_i=\max\{w_i,\tau\}$.  If $i\not\in S$, $\hat w_i=0$. We will prove
\begin{equation}\label{eq:unbias}
\Ep{\hat w_i}=w_i
\end{equation}
By linearity of expectation, if we want to estimate the total
weight of an arbitrary subset $I\subseteq [n]=\{0,1,\ldots,n-1\}$, 
we just sum the
corresponding weight estimates in the sample, that is,
\begin{equation}\label{eq:subset}
\Ep{\sum_{i\in S\cap I}\hat w_i}=\sum_{i\in I}w_i
\end{equation}
Ties between priorities happen with probability zero, and
can be resolved arbitrarily. We resolve them in favor
of earlier items. Thus we view priority $q_i$ as higher
than $q_j$, denoted $q_i\succ q_j$, if either $q_i>q_j$ or $q_i=q_j$ and
$i<j$. With any such resolution of ties, priority sampling 
works even if some weights are zero.

Note that in the case of unit weights, $\tau$ is just 
the $(k+1)^{\mathrm{th}}$ largest value $1/\alpha_i$, and 
then \req{eq:subset} simplifies to 
\begin{equation}\label{eq:unit}
\Ep{k\tau}=n.
\end{equation}
This unit case is a classic theorem in order statistics
(see e.g.,~\cite{AB88,David81}).

\subsection{Selecting Subsets}
We will now, with a few examples, illustrate how subsets could
be selected. The basic point is that an item, besides the weight,
has other associated information, and selection of an item may
be based on all its associated information. To estimate the total
weight of all selected items, we sum the weight estimates of
all sampled items that would be selected. We note that the examples
below could be based on any kind of sampling. What distinguishes
priority sampling is the quality of the answers.

\paragraph{Internet traffic analysis}
Our motivating application comes from Internet traffic
analysis. Internet routers export information about transmissions of
data passing through. These transmissions are called flows.  A flow
could be an ftp transfer of a file, an email, or some other collection
of related data moving together. A flow record is exported with
statistics such as summary information such as application type,
source and destination IP addresses, and the number of packets and
total bytes in the flow.  We think of byte size as the weight.

We want to sample flow records in such a way that we can answer
questions like how many bytes of traffic came from a given customer or
how much traffic was generated by a certain application. Both of these
questions ask what is the total weight of a certain selection
of flows.  If  we knew in advance of
measurement which selections were of interest, we could have a counter
for each selection and increment these as flows passed by.
The challenge here is that we must not be constrained to selections
known in advance of the measurements. This would preclude
exploratory studies, and would not allow a change in routine questions
to be applied retroactively to the measurements. 

A killer example where the selection is not known in advance was the
tracing of the {\em Internet Slammer worm} \cite{Slammer}. It turned
out to have a simple signature in the flow record; namely as being UDP
traffic to port 1434 with a packet size of 404 bytes.  Once this
signature was identified, the historical development of the worm could
be determined by selecting records of flows matching this signature
from a data base of sampled flow records.

We note that data streaming algorithms have been developed that
generalizes counters to provide answers to a range of selections such
as, for example, range queries in a few dimensions
\cite{Mut04}. However, each such method is still restricted to a
limited type of selection to be decided in advance of the measurements. 

\paragraph{External information in the selection} In our next example,
suppose Wallmart saved samples of all their sales where each
record contained information such as item, location, time, and price.
Based on sampled records, they might want to ask questions like
how many days of rain does it take before we get a boom in the sale
of rain gear. Knowing this would allow them to tell how long the would
need to order and disperse the gear if the weather report promissed
a long period of rain. Now, the weather information was not part
of the sales records, but if they had a data base with
historical weather information, they could look up each sampled sales
record with rain gear, and check how many days it had rained at
that location before the sale.

The important lesson from this example is that selection can
be based on external information not even imagined relevant at the
time when measurements are made. Such scenarios preclude any
kind of streaming algorithm based on selections of limitated complexity,
and shows the inherent relevance of sampling preserving full records
for the perpose of arbitrary selections.

\subsection{Relation to classic sampling schemes}
What distinguishes priority sampling is how well it does in the common
case of a heavy tailed weight distribution \cite{PKC96}. The problem
with {\em uniform sampling\/} is that it is likely to miss out on the
small proportion of heavy items.  An alternative is {\em weighted
sampling with replacement} where each sample is chosen independently,
each item being selected with probability proportional to its
weight. The problem is that we are likely to get many duplicates of
the heavy items, and hence provide less information on lighter
items. A variant of weighted sampling with replacement for integer
weights is to {\em divide them into unit weights}. This way we can get
at most $w_i$ samples of units from item $i$. However, when weights
are large compared with the number of samples, this is still very
similar to the basic weighted sampling without replacement.

The above observations suggest that we need {\em weight-sensitive sampling
  without replacement}. For example, we can perform weighted sampling
with replacement, but skip duplicates until we have the desired number
of samples.  The book \cite{BH83} mentions 50 such schemes, but none
of these provides estimates of sums. The basic problem is that the
probability that a given item is included in the sample is a
complicated function of all the involved weights.  

Clearly priority
sampling acts without replacement. To see that it is
weight-sensitive, suppose we have an item $i$ which is $r=w_j/w_i\geq 1$
times smaller than an item $j$. Then the probability that $i$ gets
higher priority than $j$ is $1/2r$. More precisely,
\[\Pr[q_i>q_j]=\Pr[w_i/\alpha_i>w_j/\alpha_j]
=\Pr[\alpha_i<\alpha_j/r]=\int_0^1 \alpha_j/r\; d
\alpha_j=1/2r\]
Priority sampling is thus weight-sensitive without
replacement, and, as stated in \req{eq:subset}, it provides simple 
unbiased estimates of arbitrary sums. 
We will present tests of priority sampling on real Internet data, and
see that estimating subset sums needs orders of magnitude fewer samples
than uniform sampling and weighted
sampling without replacement.

\subsection{Outline of the Paper}
The rest of the paper is organized as follows. In \Sref{sec:unbias} we present
the proof that priority sampling provides unbiased estimators as stated in
\req{eq:unbias}. In addition we will show how we can estimate
the variance of our subset sum estimates. This
relies on the striking property of priority
sampling that we establish, namely, 
that with more than one sample, the covariance between different
weight estimates is zero. 
In \Sref{sec:threshold} we compare our new priority sampling with threshold
sampling from \cite{DLT05}, a scheme which is very closely
related but does not provide
a specified number of samples.
In \Sref{sec:experiments}, we present experiments
with priority sampling on real data from the Internet, demonstrating
orders of magnitude gain in accuracy in estimation weight sums,
as compared with uniform sampling
and weighted sampling without replacement.
In \Sref{sec:analysis}, we analyze the performance of the different
sampling schemes in some simple cases in order to gain further understanding
of the experiments. In \Sref{sec:conj-opt}, we conjecture an extremely 
strong near-optimality; namely that for
any weight sequence, there exists no specialized scheme for
sampling $k$ items with unbiased weight estimators that gets smaller
total variance than priority sampling with $k+1$ items. This
conjecture was recently settled by Szegedy \cite{Sze05}.q
In \Sref{sec:alg}, we show how we can 
maintain a priority
sample of size $k$ for a stream of weighted items, spending only constant
time on each item as it comes by. We finish with some concluding
remarks in \Sref{sec:conclusion}.

A preliminary version of parts of this work was published in a
conference proceeding \cite{DLT04}, including the basic announcement
of the priority sampling scheme. Our original proof of
\req{eq:unbias} was based on the standard proofs for the known unit
case~\cite{AB88,David81}, but here we present a much simpler
combinatorial proof and include an entirely new analysis of variance
and covariance. The experiments reported here are all new, and so is
most of the analysis of simple cases, as well as the conjecture
concerning near-optimality.

\section{Unbiased estimation with priority sampling}\label{sec:unbias}
In this section, we will show that priority sampling yields unbiased
estimates of subset sums as stated in \req{eq:unbias}.  The proof is
simpler and more combinatorial than the standard proofs for the known
unit case~\cite{AB88,David81}. We will also show how to form unbiased
estimators of secondary weights. Finally, we consider variance
estimation. We show that there is no covariance between the weight estimates
of different items, and that we can get unbiased estimates of
the variance of any subset sum estimate.

Recall that we consider items with
positive weights $w_0,...,w_{n-1}$. For each item $i\in [n]$, we
generate an independent uniformly distributed random number $\alpha_i\in(0,1)$,
and a priority $q_i=w_i/\alpha_i$. Priority $q_i$ is higher than $q_j$,
denoted $q_i\succ q_j$, if either $q_i>q_j$, or $q_i=q_j$ and
$i<j$. A priority sample $S$ of size $k$ consists of
the $k$ items of highest
priority. The threshold $\tau$ is the $(k+1)^{\mathrm{st}}$ highest
priority. Then $i\in S\iff q_i\succ\tau$. Each $i\in S$ gets a weight
estimate $\hat w_i=\max\{w_i,\tau\}$. Also, for $i\not\in S$, we
define $\hat w_i=0$. Now \req{eq:unbias} states that $\E[\hat
w_i]=w_i$.

We will prove that \req{eq:unbias} holds for an item $i$ no
matter which values the other $\alpha_j,\ j\neq i$ take. 
Fixing these values, we fix all the other priorities $q_j,\ j\neq i$.
Let $\tau'$ be the $k^{\mathrm{th}}$
highest of these other priorities. We can now view
$\tau'$ as a fixed number. More formally, our analysis is conditioned
on the event $A(\tau')$ of $\tau'$ being the $k^{\mathrm{th}}$
highest among the priorities $q_j,\ j\neq i$,  and we will
prove 
\begin{equation}\label{eq:unbias-tau'}
\E[\hat w_i|A(\tau')]=w_i.
\end{equation}
Proving \req{eq:unbias-tau'} for any value of $\tau'$ implies \req{eq:unbias}.
The essential observation is as follows.
\begin{lemma}\label{lem:tau'} 
Conditioned on $A(\tau')$, item $i$ is picked with probability
$\min\{1,w_i/\tau'\}$, and if picked, $\tau=\tau'$.
\end{lemma}
\begin{proof}
We pick $\alpha_i\in(0,1)$ uniformly at random, thus fixing
$q_i=w_i/\alpha_i$. If $q_i\prec \tau'$, there are at least
$k$ priorities higher than $q_i$, so $i\not\in S$. Conversely, if
$q_i\succ \tau'$, then $\tau'$ becomes the $(k+1)$th priority 
among all priorities, so $\tau'=\tau$, and then $i\in S$. Finally,
\[\Prp{i\in S|A(\tau')}=\Prp{q_i\succ\tau'}=\Prp{\alpha_i<w_i/\tau'}
=\min\{1,w_i/\tau'\}\]
\end{proof}
From Lemma\tref{lem:tau'}, we get
\begin{eqnarray*}
\E[\hat w_i|A(\tau')]&=&\Prp{i\in S|A(\tau')}\times
\Ep{\hat w_i|i\in S\wedge A(\tau')}\\
&=&\min\{1,w_i/\tau'\}\times\max\{w_i,\tau'\}\\
&=& w_i
\end{eqnarray*}
The last equality follows by observing that both the $\min$ and
the $\max$ take their first, respectively their second value, depending
on whether or not $w_i\geq \tau'$. This completes
the proof of \req{eq:unbias-tau'}, hence of \req{eq:unbias}  

\subsection{Zero weight items and sampling it all}
We note here that priority sampling, as defined above, works
even in the presence of zero weights. First we
note that $w_i=0\iff q_i=w_i/\alpha_i=0$ while
$w_i>0\iff q_i=w_i/\alpha_i>w_i>0$. It follows
that zero weight items can only be sampled
if all positive weight items have been sampled. Moreover,
if we do sample a zero weight item $i$, we have
$\tau\prec q_i=w_i=0$, so $\tau=0$, and then
$\hat w_j=w_j$ for all items $j$. Having noted that zero weight
items do not cause problems, we will mostly ignore them.

Above we have assumed $k<n$, but we note a natural view of a priority
sample of everything, that is, with $k=n$. We define an $(n+1)^{th}$
priority $\tau=q_n=0$, as if we had an extra zero weight $w_n=0$.
Then $q_i\succ \tau=q_n$ for all $i\in[n]$, so all items get
sampled. Moreover $\hat w_i=\max\{w_i,\tau\}=w_i$, so the weight
estimate is equal to the original weight.

\subsection{Secondary variables}
Suppose that each item $i$ has a secondary variable $x_i$.
We can then use \req{eq:unbias} to give unbiased estimators of
corresponding secondary subset sums. More precisely,
we set $\hat x_i=\hat w_i x_i/w_i$. That is
$\hat x_i=\max\{w_i,\tau\}x_i/w_i=\max\{1,\tau/w_i\}x_i$ 
if $i$ is sampled; $0$ otherwise.
Then \req{eq:unbias} implies $\E[\hat x_i]=x_i$.

An application could be to deal with negative and positive 
weights $x_i$. We could define the priority weights as their absolute
values, that is, $w_i=|x_i|$, and use these 
non-negative weights in the priority sample. 

Another application could be if we had several different variables
for each item. Instead of making an independent priority sample for
each variable, we could construct a compromise weight. For example,
for each item, the weight could be a weighted sum of all the associated
variables.

\subsection{Variance estimation for a single item}\label{sec:item-var-est}
We now provide a simple variance estimator
\[\hat v_i=\left\{\begin{array}{ll}
w_i\tau\max\{0,\tau-w_i\}&\mbox{if }i\in S\\
0&\mbox{if }i\not\in S
\end{array}\right.\textnormal,\]
and show that it is unbiased, that is,
\begin{equation}\label{eq:var-est}
\Ep{\hat v_i}=\Var[\hat w_i].
\end{equation}
As in the proof of \req{eq:unbias}, we define $A(\tau')$ to be the
event that $\tau'$ is the $k^{\mathrm{th}}$ highest among the 
priorities $q_j,\ j\neq i$. We will prove
\begin{equation}\label{eq:var-est-tau'}
\Ep{\hat v_i|A(\tau')}=\E[\hat w_i^2|A(\tau')]-w_i^2.
\end{equation}
From Lemma\tref{lem:tau'}, we get
\begin{eqnarray*}
\E[\hat v_i|A(\tau')]&=&\Prp{i\in S|A(\tau')}\times
\Ep{\hat v_i|i\in S\wedge A(\tau')}\\
&=&\min\{1,w_i/\tau'\}\times \tau'\max\{0,\tau'-w_i\}\\
&=&\max\{0,w_i\tau'-w_i^2\}.\\
\end{eqnarray*}
On the other hand,
\begin{eqnarray*}
\E[\hat w_i^2|A(\tau')]&=&\Prp{i\in S|A(\tau')}\times
\Ep{\hat w_i^2|i\in S\wedge A(\tau')}\\
&=&\min\{1,w_i/\tau'\}\times \max\{w_i,\tau'\}^2\\
&=&\max\{w_i^2,w_i\tau'\}.\\
\end{eqnarray*}
This establishes \req{eq:var-est-tau'} and hence \req{eq:var-est}.

\subsection{Covariance}
Assuming that we sample more than one item, 
we will show that the covariance between
our weight estimates is zero , that is, for $k>1$ and $i\neq j$,
\begin{equation}\label{eq:covar}
\Ep{\hat w_i\hat w_j}=w_iw_j
\end{equation}
If $k=1$, we have $\Ep{\hat w_i\hat w_j}=0$ since we cannot sample
both $i$ and $j$.

Note that \req{eq:covar} is somewhat counter-intuitive in that if we
sample $i$ then this reduces the probability that we also sample $j$.
However, the assumption that $i$ is sampled affects the threshold
$\tau$ and thereby the weight estimate $\hat w_j$ and somehow,
the different effects cancel out.

We will prove \req{eq:covar} via the following common generalization
of \req{eq:covar} and \req{eq:unbias} holding for any 
$I\subset [n],\ |I|\leq k$:
\begin{equation}\label{eq:prod}
\Ep{\prod_{i\in I}\hat w_i}=\prod_{i\in I} w_i
\end{equation}
If $|I|>k$, we have $\Ep{\prod_{i\in I}\hat w_i}=0$ since
at most $k$ items are sampled with $\hat w_i>0$.

The proof of \req{eq:prod} generalizes that of \req{eq:unbias}.
Inductively on the size of $I$, we will prove that
\req{eq:prod} holds no matter what values all the other $\alpha_j$,
$j\not\in I$ take. The equality is trivially true in the base
case where $I=\emptyset$ and the products equals one. 

Thus, for all $j\not\in I$, fix all $\alpha_j\in(0,1)$ and 
priorities $q_j=w_j/\alpha_j$. Fix
$\tau''$ to be the $(k-|I|+1)^{\mathrm{th}}$ highest of these priorities
$q_j\ j\not\in I$. This priority exists because $k\leq |I|<n$.
Next for $i\in I$, we pick $\alpha_i\in (0,1)$ and set
$q_i=w_i/\alpha_i$. We can now have at most $(k-|I|)+|I|$ priorities
below $\tau''$, so $\tau''$ is at least as big as our
new threshold $\tau$.

Consider the case that $I$ has a weight $w_h\geq\tau''$.
Fix $\alpha_h\in(0,1)$ arbitrarily. Then
$q_h>w_j\geq \tau''\geq\tau$, so item $h$ is sampled with $\hat
w_h=\max\{w_h,\tau\}=w_h$. Hence $\Ep{\prod_{i\in I}\hat
w_i}=w_h\Ep{\prod_{i\in I\setminus\{m\}}\hat w_i}$. We have now
fixed all $\alpha_j$, $j\not\in I\setminus\{m\}$, and by
induction, $\Ep{\prod_{i\in I\setminus\{m\}}\hat w_i}=
\prod_{i\in I\setminus\{m\}} w_i$. This completes the
proof of \req{eq:prod} in the case that $w_h\geq\tau''$.

Next consider the case that all weights from $I$ are smaller than
$\tau''$. Let $q_\ell$ be the lowest priority from $I$. If
$q_\ell\prec\tau''$, then there are at least
$(k-|I|+1)+|I\setminus\{\ell\}|=k$ priorities higher than $q_\ell$, so
$q_\ell\not\in S$, and $\hat w_\ell=0=\prod_{i\in I}\hat w_i$. Thus,
if $q_\ell\prec\tau''$, there is no contribution to $\Ep{\prod_{i\in
I}\hat w_i}$.

Conversely, if $q_\ell\prec\tau''$, then all priorities from $I$ are
bigger than $\tau''$. In this case there are exactly $(k-|I|)+|I|=k$ 
priorities higher than
$\tau''$, so $\tau''$ becomes our threshold $\tau$. Then each
$i\in S$ are sampled. Since $w_i\leq\tau''=\tau$, we
get $\hat w_i=\max\{w_i,\tau\}=\tau$. Hence 
$\prod_{i\in I}\hat w_i={\tau''}^|I|$.  Since no weights in $I$ is higher
than $\tau''$, the probability that all their priorities are bigger is
$\prod_{i\in I}(w_i/\tau'')$. Thus, the contribution to
$\Ep{\prod_{i\in
I}\hat w_i}$ is ${\tau''}^|I|\prod_{i\in I}(w_i/\tau'')=\prod_{i\in I}w_i$.
This completes the proof of 
\req{eq:prod}  in the remaining case where $w_h<\tau''$. 

\subsection{Variance estimation over any subset}
We can now use our variance estimator from \Sref{sec:item-var-est}
to estimate the variance over any subset.
By \req{eq:covar} and \req{eq:var-est} we get an unbiased
estimator of the variance of any subset sum estimate simply
by summing the variance estimators from the subset, that is,
if $k>1$ for any subset $I\subseteq[n]$,
\begin{equation}\label{eq:var-sum}
\Var[\sum_{i\in S\cap I}\hat w_i]=\Ep{\sum_{i\in S\cap I}\hat v_i}
\end{equation}
In fact, \req{eq:var-sum} also holds if $k=1$, but this
is because $\Var[\sum_{i\in S\cap I}\hat w_i]=\infty$ for any
non-empty subset $I$. We shall return to this point later in 
\Sref{sec:singe-inf}.

\section{Comparison with a fixed threshold scheme}\label{sec:threshold}
It is instructive to compare our priority sampling scheme with
threshold sampling from \cite{DLT05}. In that approach each item
is sampled independently, so we do not control the exact number of
samples. Before sampling, a fixed threshold $\tau^{THR}$ is chosen.
An item $i$ is sampled if $w_i\geq\tau^{THR}$, or with probability
$w_i/\tau^{THR}$ if $w_i\leq \tau^{THR}$. We denote the set of
selected items by $S^{THR}$.
 
To see the relation to priority
sampling, note that threshold sampling can be expressed in a manner
similar to priority sampling as follows: 
generate a random number $\alpha_i\in(0,1]$ and sample item $i$
if $q_i=w_i/\alpha_i>\tau^{THR}$.
As in our new scheme, the sampled items get weight estimate
$\hat w_i^{THR}=\max\{w_i,\tau^{THR}\}$ whereas $\hat w_i^{THR}=0$ if $i\not\in S^{THR}$.
Thus the only difference between priority sampling and the
threshold sampling from \cite{DLT05} is in the choice of the 
threshold. In threshold sampling, 
the threshold is fixed independent of the random choices. Thus the threshold
determines only the expected number of independent samples, not the
actual random number of samples. By contrast, in priority sampling,
the threshold is picked depending on
the random choices so as to get a fixed number of dependent samples.
We note that it is far from obvious that such a threshold could be
chosen without violating the unbiasedness of estimation.

\subsection{Optimality of the fixed threshold scheme}\label{sec:thr-opt}
In \cite{DLT05}, the fixed threshold approach to independent
sampling is proved to give an optimal trade-off between variance and
sampling rate. More for an item $i$
with weight $w_i$, we have to decide on a sampling probability $p_i$.
If $i$ is not picked, the weight estimate is zero, that is, 
$\hat w_i(p_i)=0$. To get an unbiased estimator, if item $i$ is picked, 
it should have weight estimate $\hat w_i(p_i)=w_i/p_i$. 
Then $\Ep{\hat w_i(p_i)}=w_i$. Generally, we want to
sample few items, yet keep the variance low. This motivates an
objective of the form
\[\mbox{minimize } p_i+\beta\,\Varp{\hat w_i(p_i)}\]
Here
\[\Varp{\hat w_i(p_i)}=\Ep{(\hat w_i(p_i))^2}-w_i^2\mbox{ where }
\Ep{(\hat w_i(p_i))^2}=p_i(w_i/p_i)^2=w_i^2/p_i.\]
Thus we want to 
\[\mbox{minimize } p_i+\beta\,w_i^2/p_i\]
For $p_i\in[0,1]$, the solution is to
set $p_i=\min\{1,\sqrt\beta w_i\}$. With
$\sqrt\beta=1/\tau^{THR}$ this is equivalent to the fixed threshold scheme.
That is, for any choice of $\tau^{THR}$, the fixed threshold scheme picks 
the $p_i=\max\{1,w_i/\tau^{THR}\}$ so as to
\begin{equation}\label{eq:thr-ind-opt}
\mbox{minimize } p_i+(1/\tau^{THR})^2\Varp{\hat w_i(p_i)}.
\end{equation}
Summing over the whole stream of items, we
\begin{equation}\label{eq:thr-tot-opt}
\mbox{minimize } \sum_{i\in[n]}\left(p_i+(1/\tau^{THR})^2
\Varp{\hat w_i(p_i)}\right)
=\sum_{i\in[n]}p_i+1/(\tau^{THR})^2\Varp{\sum_{i\in[n]}\hat w_i(p_i)}
\end{equation}
We now constrain ourselves to getting an expected
number $k$ of samples. To minimize the total variance,
we just have to identify $\tau^{THR}$ such that
\[\sum_{i\in[n]}{p_i}=\sum_{i\in[n]}\min\{1,w_i/\tau^{THR}\}=k\]
With this value of $\tau^{THR}$, the fixed threshold scheme from
\cite{DLT05} minimizes the total variance subject to unbiased
estimation and an expected number $k$ of samples. Any other
assignments of individual sampling probabilities $p_i$ will do
worse. The quality of our new scheme is largely inherited from this
fixed threshold scheme, but we have some extra variability due to the
variability of the threshold.

\section{Experiments}\label{sec:experiments}
We tested priority sampling on 10 minutes of flows from an Internet
gateway router. For increasing sample sizes, we wanted to check our
ability to estimate subset sums where each subset was defined by flows
originated by certain applications such as FTP and web traffic. This
illustrates how priority sampling can be used today in a 
backbone network.  The basic flow statistics for the different applications is
presented in Table\tref{tab:stats}.
\begin{table}
\begin{center}
\begin{tabular}{|lrrrrrrr|}\hline
application & bytes & \% of traffic & \# flows & \% flows & max flow size &
average & min\\\hline
all & 4265677642 & 100.00 & 85680 & 100.00 & 3372865057 & 49786 & 28 \\
ftp & 3394832734 & 79.58 & 727 & 0.84 &3372865057 & 4669646  & 40 \\
web & 80120429 & 1.87 & 7787 & 9.08 & 3139196 & 10289 & 40 \\
dns & 4083277 & 0.09 & 40767 & 47.58 & 621812 & 100  & 40 \\
\hline
\end{tabular}
\caption{Statistics on ten minutes of flows from an Internet gateway 
router showing traffic from some different applications. Note that nearly
half the flows belong to applications not mentioned.}\label{tab:stats}
\end{center}
\end{table}
We compared the following sampling schemes:
\begin{description}
\item[PRI] our new priority sampling.
\item[U$-$R] uniform sampling without replacement.
\item[W$+$R] weighted sampling with replacement.
\item[THR] the fixed threshold sampling from \cite{DLT05}
as described in \Sref{sec:threshold}.
\end{description}
For weighted sampling with replacement, we note that
there are two alternative ways of deriving weight estimates.
More precisely, we have a list $S^{W+R}$ of $k$ samples. Each
sample $S^{W+R}[j]$ is independent and equals item $i$ with
probability $w_i/W$ where $W$ is the total weight. The simplest
unbiased estimator of $w_i$ counts duplicates, estimating
$w_i$ as $|\{j|S^{W+R}[j]\}|W/k$. However, we get a smaller
variance if we just consider whether item $i$ is present
in $S^{W+R}$. The probability of this event is
\[p_i^{W+R}=1-(1-w_i/W)^k,\]
and then we get the unbiased weight estimator:
\[\hat w_i^{W+R}=\left\{\begin{array}{ll}
w_i/p_i^{W+R}&\mbox{if }i\in S^{W+R}\\
0&\mbox{otherwise}
\end{array}\right.\]

\begin{figure}
\begin{center}
\begin{tabular}{cc}
\epsfig{file=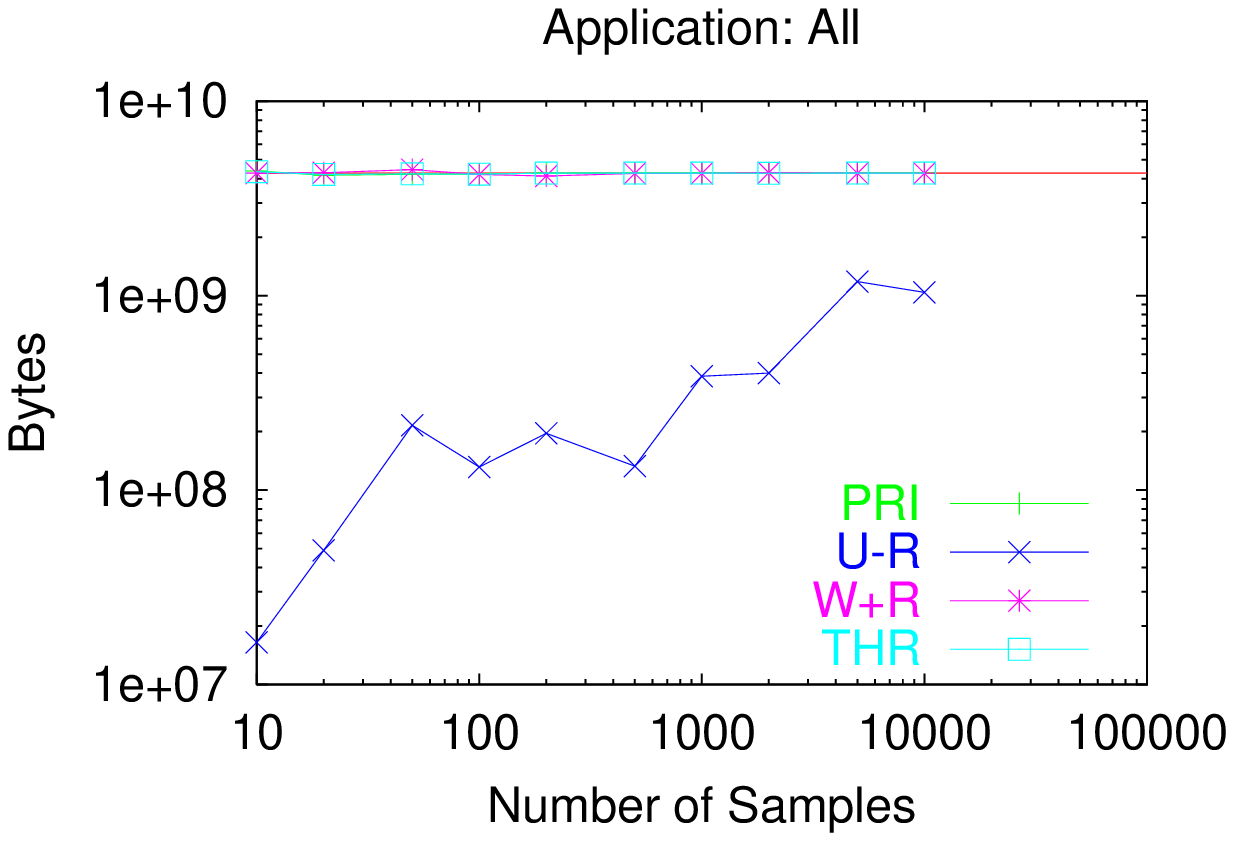,width=3.4in}&\kern -30pt
\epsfig{file=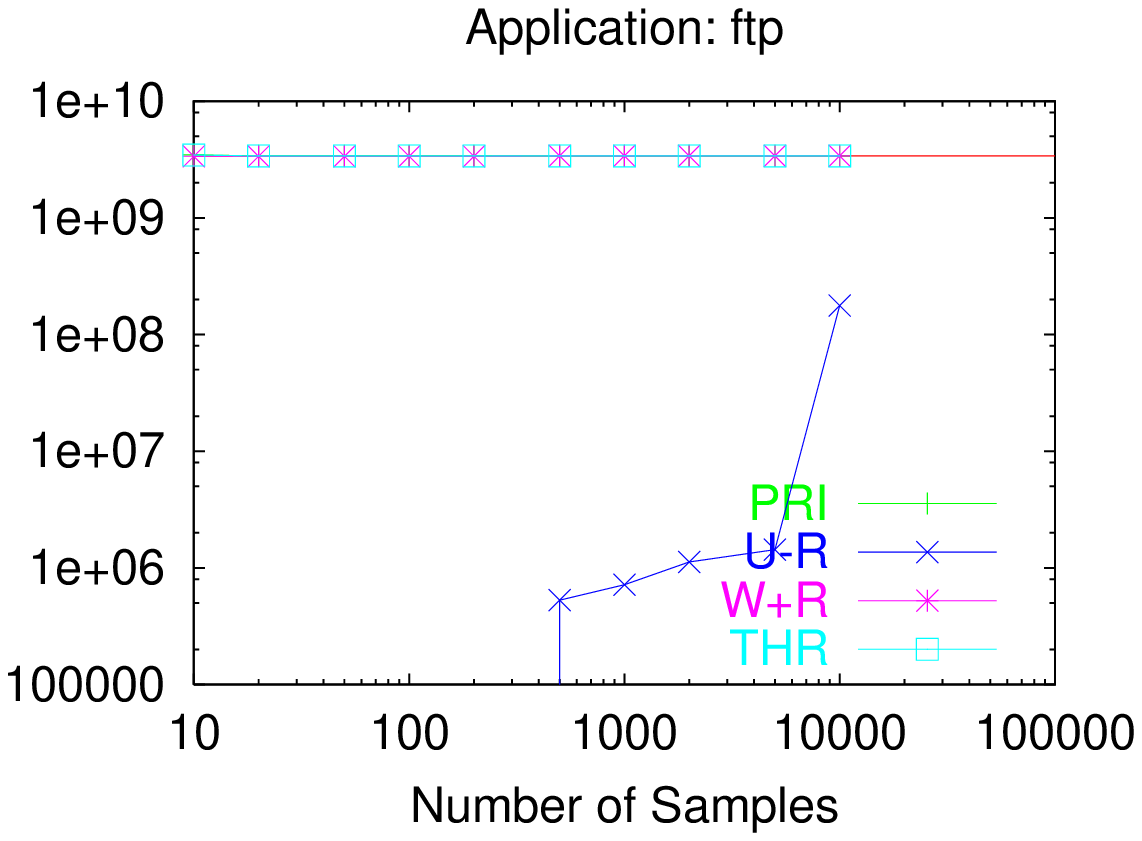,width=3.4in}\\
\epsfig{file=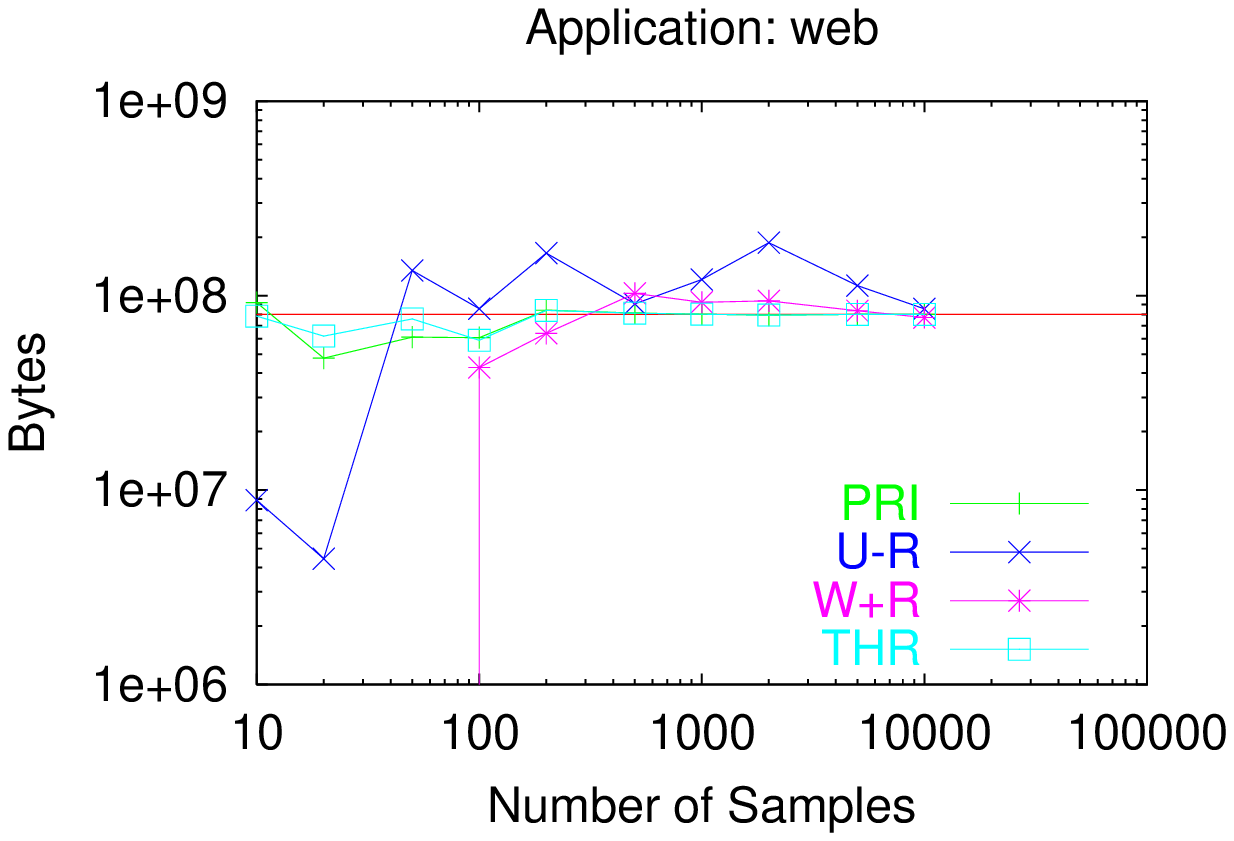,width=3.4in}&\kern -30pt
\epsfig{file=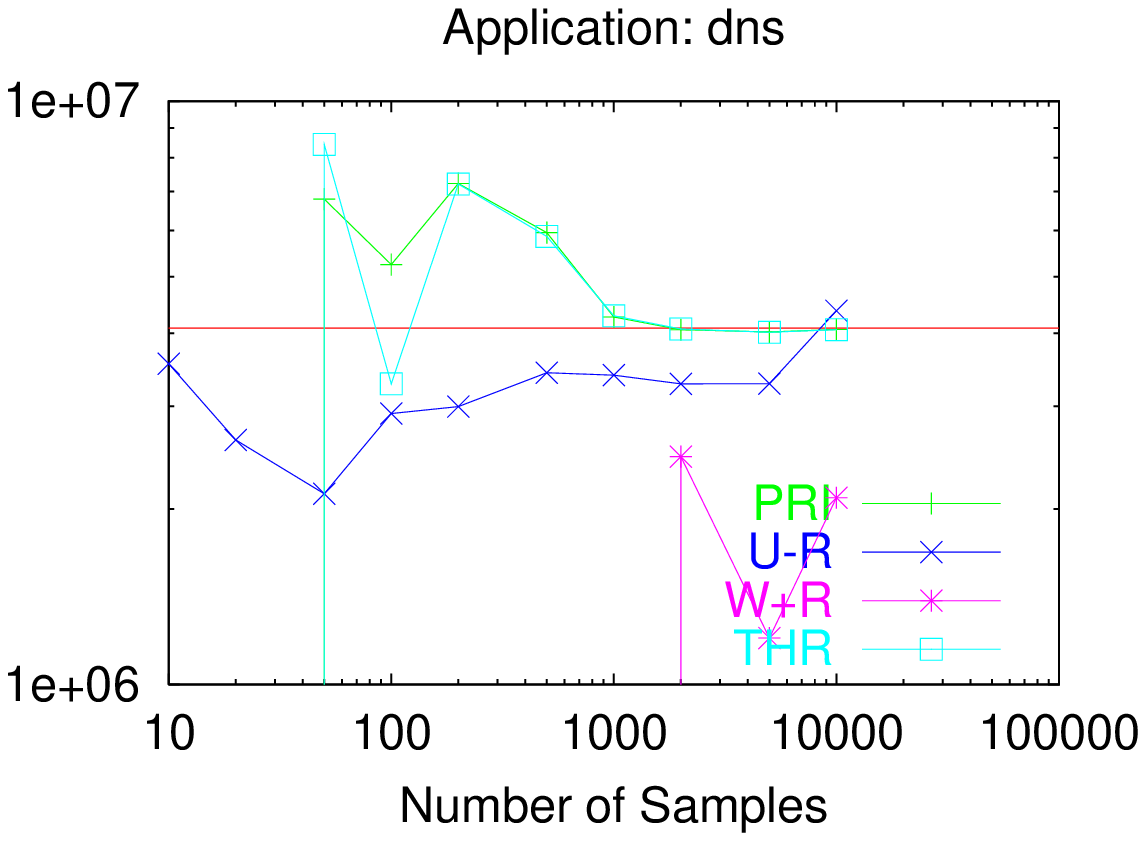,width=3.4in}\\
\end{tabular}
\end{center}
\caption{Estimating traffic from different applications with
different sampling strategies. The red line shows the
true traffic from each application.}\label{fig:applications}
\end{figure}

We now describe the setup of the experiments; the interpretation of
the results follows in the next sections.
In Figure\tref{fig:applications} we compare the estimation accuracy of
the different sampling schemes on the data summarized in 
Table\tref{tab:stats}.
For each sampling scheme, we progressively increased the size of 
sample by selecting more items from the data, estimating total weight
in each application subset of the total flows for each sample size.

\begin{figure}
\centerline{\leavevmode\epsfig{file=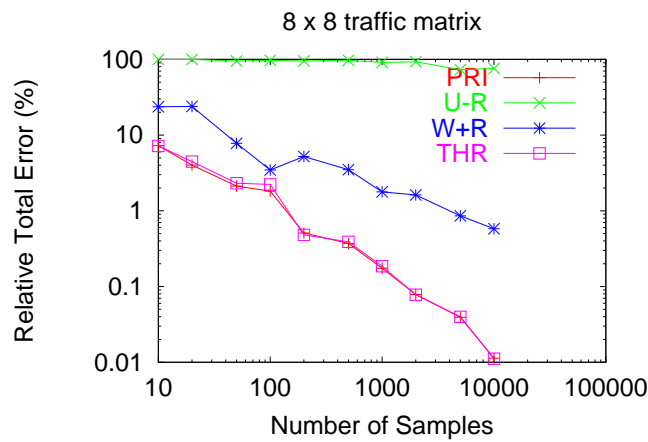,width=3.4in}}
\caption{Estimating a traffic matrix with different sampling strategies.
We divide the total error over all entries with the total 
traffic.}\label{fig:matrix}
\end{figure}

In Figure\tref{fig:matrix}, the same samples are used to estimate an
$8\times 8 =64$ entry traffic matrix.
Each matrix element corresponds to the traffic between an input and
output interface on a router. We estimate the total bytes for each matrix element. Our
accuracy measure is average over all elements of the relative
estimation error.
For priority sampling (PRI), uniform sampling 
without replacement (U$-$R), and weighted sampling with replacement 
(W$+$R), the number $k$ of samples is exact. 

In threshold sampling (THR), the threshold determines only the
expected number of samples. 
For each item $i$, we used the same priority $q_i=w_i/\alpha_i$
for priority sampling and threshold sampling. In priority sampling, 
we picked exactly $k$ samples using the $(k+1)^{th}$ priority $\tau$ as a 
threshold. In threshold sampling, we computed the threshold $\tau^{THR}$ 
giving an expected number $k$ of samples. Thus, for a given $k$,
the only difference is in the choice of threshold.

\begin{figure}
\centerline{\leavevmode\epsfig{file=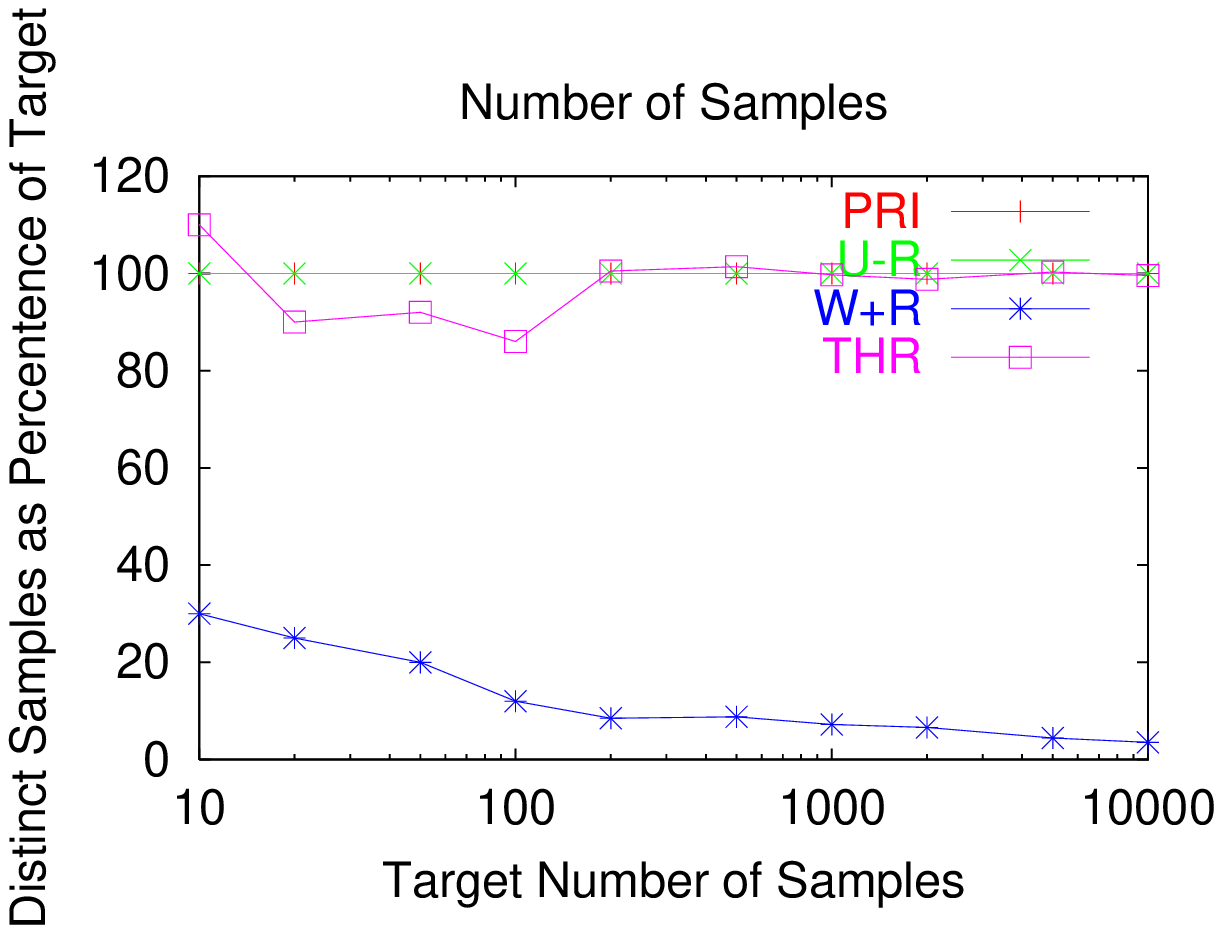,width=3.4in}}
\caption{Number of distinct samples as percentage of target $k$. 
}\label{fig:distinct-samples}
\end{figure}

Finally, Figure\tref{fig:distinct-samples} tells the number of
distinct samples as a percentage of the target. For
priority sampling (PRI) and uniform sampling (U$-$R) we have no replacement,
so we get exactly $k$ distinct samples, that is, 100\%. With
weighted sampling with replacements (W$+$R) the duplicates mean that
we get less distinct samples. Finally, with threshold sampling (THR),
all samples are distinct, but we only have an expected number $k$ of samples, hence the deviation from the target $k$.

\subsection{Discussion}
The quality of a sampling scheme is the number of samples it
takes before the estimates converges towards the true value.

\subsubsection{Sampling exactly $k$ samples}
First we compare our priority sampling (PRI) scheme with the other schemes
providing an exact number $k$ of samples, that is, with uniform sampling
without replacement (U$-$R) and weighted sampling without replacement
(W$+$R). In \Fref{fig:applications} and \tref{fig:matrix}, we 
see that priority sampling provides very substantial gains in accuracy
over
the other schemes.

When comparing the curves, there are two points to consider. 
One is how many samples it takes before we get one 
from a given application. This is the point at which we 
get our first non-zero estimates. Second we consider
how quickly we converge after this point.

\paragraph{Number of samples needed to hit an application} With uniform
sampling, the number of samples expected before we get
one from a given application is  roughly the total number of flows
divided by the number of application flows. In that regard,
ftp traffic is clearly the worst.

With weighted sampling without replacement, the expected number is  
roughly the total traffic divided by the application traffic. The worst
application here is dns traffic which was the best application for uniform 
sampling.

Priority sampling is like weighted sampling without replacement but it
avoids making duplicates of dominant items. If the dominant items
are outside the application, we waste at most one sample on each.
The impact is clear for dns traffic where we get the first sample about
30 times earlier with priority sampling than we did with weighted sampling
without replacement. A more direct illustration of the problem is
found in \Fref{fig:distinct-samples} where we see how the fraction
of distinct samples drops in weighted sampling without replacement.

\paragraph{Convergence after first hitting an application} After we have started getting
samples from an application, uniform sampling may still have problems
with convergence. This typically occurs if the weight distribution
within the application is heavy-tailed. Once again, ftp traffic is the
worst application, this time because it has a dominant flow with more
than 99\% of its traffic. Until this flow is sampled, we expect to
underestimate. If it is sampled early, we will hugely overestimate,
although this is unlikely.  The typical heavy-tail behavior is that
the estimate grows as we catch up with more and more dominant items.
We see this phenomena both for ftp traffic and for all traffic
combined.

With weighted sampling without replacement and with priority sampling,
we get quicker convergence as soon as we start having samples
from an application. Neither scheme has any problems with
skewed weight distributions within the applications. For example,
we see that weighted sampling without replacement starts slower than
uniform on web traffic, yet it ends up converging faster. Similarly,
priority sampling starts slower than
uniform on dns traffic, yet it converging faster. 

\paragraph{The traffic matrix}
\Fref{fig:matrix} shows the average relative error over 
$8\times 8=64$ entries. We note
first the poor performance of uniform sampling. In fact, it is
only luck that the error with uniform is remains below 100\%. This
is because we miss the dominant items and get under-estimates
that can never be by more than 100\%. We could instead have
gotten a dominant item early, leading to a huge over-estimate
by far more than 100\%. 

Comparing priority sampling with weighted sampling without replacement 
the faster convergence of priority sampling is very clear. For example,
priority sampling gets down around a 1\% error with about 150 samples
whereas weighted sampling with replacement needs about 3000 samples,
and the weighted sampling falls further behind with smaller errors
because it gets more and more duplicates.

\subsection{Priority sampling versus threshold sampling}
A reason to believe that priority sampling works very well
for a fixed number of samples is its similarity with threshold
sampling which for an expected number of independent samples
minimized the total variance. In Figure\tref{fig:applications} 
and\tref{fig:matrix} we see that indeed priority sampling (PRI) and 
threshold sampling (THR) are
very close; neither having a systematic advantage.
Hence, in our experiment, we see now loss in quality 
going from an expected number
of samples (THR) to an exact number of samples (PRI).
The variation in the actual number of samples with THR 
shown in \Fref{fig:distinct-samples}.

As we shall see below, there are certain boundary phenomena that
makes priority sampling perform significantly worse than
threshold sampling.

\section{Analytic comparison of variance in some simple cases}
\label{sec:analysis}
In this section, we will compare the different
sampling schemes on some simple cases where
we can analyze the variance, so as to gain some intuition
for what is going on. 

Generalizing notation from \Sref{sec:threshold}, if $w$ is a weight
and $p\in[0,1]$ a sampling probability, we let $\hat w(p)$ denote the
random variable that is $w/p$ with probability $p$; $0$ otherwise.
Then 
\begin{eqnarray*}
\Ep{\hat w(p)}&=&w\\
\Ep{(\hat w(p))^2}&=&p (w/p)^2=w^2/p\\
\Varp{\hat w(p)}&=&w^2/p-w^2=w^2 \frac{1-p}p
\end{eqnarray*}
It is also convenient to define the function
\[v(w,\tau)=w\max\{0,\tau-w\}\]
Then, with fixed threshold $\tau^{THR}$, the
variance for item $i$ is 
\begin{eqnarray*}
\Var[\hat w_i^{THR}]&=&\Varp{\hat w_i(\max\{1,w_i/\tau^{THR}\})}\\
&=& w_i^2(1/\max\{1,w_i/\tau^{THR}\}-1)\\
&=&w_i\max\{0,\tau-w_i)\\
&=&v(w_i,\tau^{THR})
\end{eqnarray*}
With our new priority sampling, the threshold changes, and
the variance of item $i$ is 
\begin{equation}\label{eq:freq}
\Varp{\hat w_i}=\int_{\tau'=0}^\infty 
f(\tau')v(w_i,\tau')\,d\tau'
\end{equation}
where $f(\tau')$ is the probability density function for $\tau'$ to be
the
$k^{th}$ threshold amongst the items $j\neq i$. With $\tau'$ thus
defined, by Lemma\tref{lem:tau'}, item $i$ is picked if
$q_i=w_i/\alpha_i>\tau'$ with $\hat w_i=\tau'$; $0$ otherwise. This
imitates the fixed threshold scheme with $\tau'=\tau^{THR}$. Thus
\req{eq:freq} follows from the previous calculation with a fixed
threshold.

Sometimes it is easier with a more direct calculation.
Summing over all $j\neq i$, we integrate over choices
of $\alpha_j$, multiply with the probability
that $q_j=w_j/\alpha_j$ is the $k$th highest priority from 
$[n]\setminus\{i\}$, 
and multiply with the variance $v(w_i,q_j)$.
That is, 
\begin{equation}\label{eq:var-expression}
\Varp{\hat w_i}=\sum_{j\in[n]\setminus\{i\}}\int_0^1
\Prp{|\{h\in [n]\setminus\{i,j\}|q_h\succ q_j\}|=k-1}v(w_i,q_j)\; d\alpha_j
\end{equation}

\subsection{Infinite variance with single priority sample}\label{sec:singe-inf}
We will show that if we only make a 
single priority sample with $k=1$, then the variance of any weight 
estimate is infinite. The proof is based on \req{eq:var-expression}.
We assume $i=0$. For a lower-bound, in the sum, we only need to consider 
one other
item $j=1$. Also, when integrating over $\alpha_1$, we only
consider very small values of $\alpha_1$. More precisely,
define $\eps=w_1/(2W)$ where $W$ is the sum of all weights. 
If $\alpha_1<\eps$, we have $q_1=w_1/\alpha_1>2W$, and then 
\begin{eqnarray*}
\Prp{|\{h\in [n]\setminus\{i,j\}|q_h\succ q_j\}|=k-1}
&=&\Prp{|\{h\in \{2,...,n-1\}|q_h\succ q_1\}|=0}\\
&>&1-\sum_{h\in \{2,...,n-1\}}\Prp{q_h>2W}\\
&=&1-\sum_{h\in \{2,...,n-1\}}(w_h/2W)\\
&>&1/2
\end{eqnarray*}
Moreover 
\[v(w_i,q_j)=v(w_0,q_1)=w_0\max\{0,w_1/\alpha_1-w_0\}>w_1/(2\alpha_1)\]
Thus, by \req{eq:var-expression}, we have
\[\Varp{\hat w_0}>\int_0^{\eps}1/2\cdot w_1/(2\alpha_1)\; d\alpha_1
=\infty\]
We note that none of the other sampling schemes considered
can get infinite variance. 

Next, we argue that the variance is bounded if we make at least
two priority samples. Again, we focus on the variance for item $i=0$.
Also, it suffices to show that the integral in \req{eq:var-expression}
is finite for each value of $j$, that is, we
want to show that
\[V_{i,j}=\int_0^1
\Prp{|\{h\in [n]\setminus\{i,j\}|q_h\succ q_j\}|=k-1}v(w_i,q_j)\; d\alpha_j\]
is bounded. Now, for $k\geq 2$,
\begin{eqnarray*}
\Prp{|\{h\in [n]\setminus\{i,j\}|q_h\succ q_j\}|=k-1}
&\leq&\Prp{|\{h\in [n]\setminus\{i,j\}|q_h\succ q_j\}|\geq 1}\\
&\leq&\sum_{h\in [n]\setminus\{i,j\}}\Prp{q_h\succ q_j}\\
&\leq&\sum_{h\in [n]\setminus\{i,j\}}\Prp{w_h/\alpha_h>w_j/\alpha_j}\\
&=&\sum_{h\in [n]\setminus\{i,j\}}\min\{1,w_h\alpha_j/w_j\}\\
&\leq&\sum_{h\in [n]\setminus\{i,j\}}(w_h\alpha_j/w_j)\\
&<&W\alpha_j/w_j
\end{eqnarray*}
Moreover, 
\[v(w_i,q_j)=w_i\max\{0,w_j/\alpha_j-w_i\}\leq w_iw_j/\alpha_j\textnormal,\]
so we get that 
\[V_{i,j}<\int_0^1 W\alpha_j/w_j\cdot w_iw_j/\alpha_j\,d\alpha_j=
\int_0^1 Ww_i\;d\alpha_j=Ww_i.\]
Hence
\[\Varp{\hat w_i}=\sum_{j\in[n]\setminus\{i\}}V_{i,j}<
n\,Ww_i,\]
so indeed the variance is bounded. Since the covariance is zero, 
it also follows that estimates of weights
of subsets are bounded. Thus we have proved
\begin{proposition} If we make a single priority sample, then
all weight estimates have infinite variance. With more
than one priority samples, all weight estimates are finite.
\end{proposition}
By contrast, with all the other sampling schemes, the variance estimates
are finite as soon as we make at least one sample.

\subsection{Unit weights}
We will now study identical unit weights, focusing on the first
item $i=0$. We will compute the exact variance for each of the sampling
schemes considered.
\paragraph{Uniform sampling without replacement}
For uniform sampling without replacement, item $0$ is picked
with probability $p_0^{U-R}=k/n$, hence with 
\[\Varp{{\hat w_0^{U-R}}}=\frac{1-p_0^{U-R}}{p_0^{U-R}}=\frac {n-k}k\]
\paragraph{Weighted sampling without replacement}
For weighted sampling with replacement, item $0$ is picked
with probability $p_0^{W+R}=1-(1-1/n)^k$, hence with 
\[\Varp{{\hat w_0^{W+R}}}=
\frac{1-p_0^{W+R}}{p_0^{W+R}}=\frac{(1-1/n)^k}{1-(1-1/n)^k}\]
For $k\ll n$, the variance approaches $\frac {n-k}k$ from above. However,
for $k=n$, the variance approaches $1/(e(1-e^{-1})=0.58..$.

\paragraph{Fixed threshold}
In the fixed threshold scheme from \cite{DLT05}, 
we set $\tau^{THR}=n/k$. Then
\begin{equation}\label{eq:var-fix-unit}
\Varp{\hat w_0^{THR}}=v_0(\tau^{THR})=w_1\max\{0,\tau^{THR}-w_i\}
=\frac{n-k}k
\end{equation}
\paragraph{Priority sampling}
For priority sampling, we will evaluate \req{eq:var-expression} exactly.
We use that 
\[\Prp{q_h\succ q_1}=\Prp{\alpha_h<\alpha_1}=\alpha_1\]
and 
\[v_0(q_1)=w_0\max\{0,q_1-w_0\}=(1/\alpha_1-1)\]
so
\begin{eqnarray*}\label{eq:var-unit}
\Varp{\hat w_0}&=&\sum_{j=1}^{n-1}\int_0^1
\Prp{|\{h\in \{2,...,n-1\}|q_h\succ q_1\}|=k-1}v_0(q_1)\; d\alpha_1\\
&=&(n-1)\int_{\alpha_1=0}^1\Prp{
B(n-1,\alpha)=k-1} (1/\alpha_1-1)\; d\alpha_1\\
&=&(n-1)\int_{\alpha_1=0}^1{n-1 \choose k-1}\alpha_1^{k-2}(1-\alpha_1)^{n-k+1}\; 
d\alpha_1\\
&=&\frac{n-k}{k-1}
\end{eqnarray*}
\paragraph{Discussion}
For unit weights, uniform sampling without replacement and threshold
sampling gets the sample variance on single item weight estimates;
namely $\frac{n-k}k$. When $k$ is not too small, priority sampling
gets nearly the same variance; namely $\frac{n-k}{k-1}$. Weighted
sampling with replacement starts doing well, but gets worse and
worse as $k$ grows. In particular, for any 
$k\geq n$, it has positive variance while all the other schemes have 
zero variance since they have no replacement.

\subsection{Large and small weights}
In this section we illustrate what happens when
different weights are involved. We consider the case where we have
$\ell$ large weights of weight $N$ and $n$ 
unit weights. The large weights are first, that
is, $w_0=\cdots=w_{\ell-1}=N$ while 
$w_\ell=\cdots=w_{n+\ell-1}=1$. We let $W=\ell N+n$ denote
the total weight. We view 
$\ell$, $n$, and $N$ as unbounded. We assume $\ell\ll n\ll \sqrt N$
and that $k\ll n$. These assumptions will help simplifying the
analysis. We will use $w_0$ as a representative for the large items
and $w_n$ as a representative for the small items. The results
variances from the different sampling schemes will be accumulated
in Table \ref{tab:big-small}.

\paragraph{Uniform sampling without replacement}
For uniform sampling without replacement, the large item $0$ is picked
with probability $p_0^{U-R}=k/(n+\ell)$, hence with 
\[\Varp{{\hat w_0^{U-R}}}=N^2\frac{1-p_0^{U-R}}{p_0^{U-R}}=N^2\frac {n+\ell-k}k
\approx N^2\frac nk.\]
For small item $n$, we have the same sampling probability, 
$p_n^{U-R}=k/(n+\ell)$, so we get
\[\Varp{{\hat w_n^{U-R}}}=\frac{1-p_n^{U-R}}{p_n^{U-R}}\approx \frac{n}{k}.\]

\paragraph{Weighted sampling with replacement}
For weighted sampling with replacement, the large item $0$ is picked
with probability $p_0^{W+R}=1-(1-N/W)^k\approx 1-e^{-k/\ell}$
hence with 
\[\Varp{\hat w_n^{W+R}}=N^2\frac{1-p_0^{W+R}}{p_0^{W+R}}
\approx N^2\frac{e^{-k/\ell}}{1-e^{-k/\ell}}=N^2/(e^{k/\ell}-1).\]
In particular, this is $\Theta(N^2)$ for $k=\Theta(\ell)$.
\footnote{$f(n)=\Theta(g(n))$ iff there exist $a,b>0$ such that
$a f(n)<g(n)<b f(n)$ for all sufficiently large $n$.}
Yet
it saves a factor $n$ over uniform sampling with replacement in
the case of large weights.

For weighted sampling with replacement, the small item $i=n$ is picked
with probability $p_n^{W+R}=1-(1-1/W)^k\approx k/W\approx k/(\ell N)\ll1$, 
hence with 
\[\Varp{\hat w_n^{W+R}}=\frac{1-p_n^{W+R}}{p_n^{W+R}}
\approx \ell N/k\]

\paragraph{Fixed threshold}
For the
fixed threshold scheme, if $k\leq\ell$, we set 
$\tau^{THR}=W/k>N$. Then for heavy item $0$,
\[\Varp{\hat w_0^{THR}}=v(w_0,\tau^{THR})=N(W/k-N)\approx N^2\frac{\ell-k}k\]
while for a light item $n$, it is 
\[\Varp{\hat w_n^{THR}}=v(w_n, \tau^{THR})=(W/k-1)\approx N\ell/k\]
On the other hand, for $k>\ell$, we 
pick a threshold below $N$; namely $\tau^{THR}=(n-\ell)/(k-\ell)$. 
Then for heavy item $0$, 
\[\Varp{\hat w_0^{THR}}=0\]
while for a light item $n$, it is 
\[\Varp{\hat w_n^{THR}}=v(w_n,\tau^{THR})=(n-\ell)/(k-\ell)\approx n/(k-\ell)\]

\paragraph{Priority sampling}
First we consider big item $0$. To compute the
variance, we sum over the events $A(m)$ that we have $m$ small items
with priorities bigger than $N$, multiplying the probability of $A(m)$
with 
\[\Ep{\hat w_0^2|A(m)}-w_0^2=\Ep{\hat w_0^2|A(m)}-\Ep{\hat w_0|A(m)}^2=
\Varp{\hat w_0|A(m)}.\]
Trivially,
$\Prp{A(m)}=\Prp{B(n,1/N)=m}$. Consider a small item $i$. Conditioned on 
having a big priority $q_i>N$, item $i$ acts like a heavy item. 
Conversely, conditioned on having a small priority $q_i<N$, item
$i$ has no impact on the weight estimate of a heavy items. Thus, in
the event $A(m)$, the variance of item $0$ is as if we had
$\ell+m$ heavy items and no small items. If $\ell+m\leq k$, the threshold
is at most $N$, and then there is no variance. If $\ell+m> k$,
the analysis from the uniform unit case shows that 
\[\Varp{\hat w_0|A(m)}=N^2\frac{\ell+m-k}{k-1}\]
Thus
\[\Varp{\hat w_0}=\sum_{m=0}^n \Prp{A(m)}\Varp{\hat w_0|A(m)}
=\sum_{m=\max\{0,k-\ell+1\}}^n\Prp{B(n,1/N)=m}N^2\frac{\ell+m-k}{k-1}\]
Since $N\gg n^2$, the first term dominates, so with
$m=\max\{0,k-\ell+1\}$, we get that
\[\Varp{\hat w_0}\approx\Prp{B(n,1/N)=m}N^2\frac{\ell+m-k}{k-1}\]
If $k<\ell$, we get $m=0$,  and then 
\[\Varp{\hat w_0}\approx \Prp{B(n,1/N)=0}N^2\frac{\ell-k}{k-1}
\approx N^2\frac{\ell-k}{k-1}\]
If $k\geq \ell$, we get $m=k-\ell+1$,  and then 
\begin{eqnarray*}
\Varp{\hat w_0}&\approx&\Prp{B(n,1/N)=k-\ell+1}N^2/k\\
&\leq&(n/N)^{k-\ell+1}N^2/k\\
&=&nN(n/N)^{k-\ell}/k
\end{eqnarray*}

We now consider the light item $n$. We are going
to prove that $\Varp{\hat w_n}\approx N\ell/(k-1)$ if $k\leq \ell$,
$\Varp{\hat w_n}\approx n\ln N$ if $k=\ell+1$, and
$\Varp{\hat w_n}\approx n/(k-\ell-1)$ if $k>\ell+1$.

We consider two different
contributions to the variance depending on whether the threshold
$\tau$ is greater than $N$. If $\tau>N$, we further distinguish
depending on whether $q_n>N$. If $\tau>N$ and $q_n\leq N$,
then $\hat w_n=0$ so the variance relative to $w_n$ is $1$. The
probability of this event is 
\[\Prp{q_n\leq N}\sum_{m=\max\{0,k-\ell+1\}}^{n-1}\Prp{B(n-1,1/N)=m}
\approx \Prp{B(n-1,1/N)=\max\{0,k-\ell+1\}}\]
If $k<\ell$, this is a variance contribution close to $1$, and if
$k\geq \ell$, the variance contribution is bounded by
$\Prp{B(n-1,1/N)=k-\ell+1}<(n/N)^{k-\ell+1}$. In either case,
this contribution to the variance is not significant.

Next consider the case that $\tau>N$ and $q_n>N$. The
probability that $q_n>N$ is $1/N$. Let $A'(m)$ denote
that event that we have $m$ small items $i\neq n$ with
$q_i>N$. Conditioned on $q_n>N$, we have $\tau>N$ if and only if
$m\geq k-\ell$. In this case, the variance contribution is
$\Ep{\hat w_n^2}-1$. However, $\hat w_n$ behaves like the
weight estimate of heavy item among $\ell+m+1$ heavy items,
so $\Ep{\hat w_n^2|q_n>N\wedge A'(m)}=N^2\frac{\ell+m+1}{k-1}$.
Thus we get a
variance contribution of
\begin{eqnarray*}
&&\Prp{q_n>N}\sum_{m=\max\{0,k-\ell\}}^{n-1}\Prp{B(n-1,1/N)=m}
(N^2\frac{\ell+i+1}{k-1}-1)\\
&\approx&1/N\cdot \Prp{B(n-1,1/N)=\max\{0,k-\ell\}}
N^2\frac{\ell+\max\{0,k-\ell\}}{k-1}\\
\end{eqnarray*}
For $k\leq \ell$, this is approximately $N\ell/(k-1)$, which
dominates the variance. For $k>\ell$, this variance contribution
is approximately,
$N\Prp{B(n-1,1/N)=k-\ell}<(n/N)^{k-\ell-1}$, which is insignificant.

We now consider the case where 
$\tau\leq N$. This requires $k>\ell$ and
is like the unit case, except that we only sample
$k'=k-\ell$ items. Hence we can apply the integral from the unit weight case,
but with the restriction that $\alpha\geq 1/N$. We then get
a variance contribution of 
\begin{eqnarray*}
&=&(n-1)\int_{\alpha=1/N}^1\Prp{ B(n-1,\alpha)=k'-1} (1/\alpha-1)\;
d\alpha\\ 
&=&(n-1)\int_{\alpha=1/N}^1{n-1 \choose
k'-1}\alpha^{k'-2}(1-\alpha)^{n-k'+1}\; d\alpha\\
\end{eqnarray*}
For $k'\geq 2$, the impact of starting the integral at $1/N$ is insignificant,
so we get an variance contribution which is approximately
$\frac{n-k'}{k'-1}=\frac{n-k+\ell}{k-\ell-1}\approx\frac{n}{k'-1}$. 
For $k'=1$, we get a
variance contribution of
\[(n-1)\int_{\alpha=1/N}^1{n-1 \choose
k'-1}\alpha^{k'-2}(1-\alpha)^{n-k'+1}\; d\alpha
<n\int_{\alpha=1/N}^1\alpha^{-1}\; d\alpha=n\ln N.\]
This completes the analysis of priority sampling for large and small weights.
A comparison of all the sampling schemes is summarized in
Table \ref{tab:big-small}.
\begin{table}
\begin{center}
\begin{tabular}{|c|c|c|c|c|}\hline
          & $1\leq k<\ell$ & $k=\ell$ & $k=\ell+1$ & $k>\ell$\\\hline
\multicolumn{5}{|c|}{large item}\\\hline
U$-$R       &  \multicolumn{4}{c|}{$N^2n/k$}  \\\hline
W$+$R       &  \multicolumn{4}{c|}{$N^2/(e^{k/\ell}-1)$}  \\\hline
THR    &  $N^2\frac{\ell-k}k$ &  $nN/\ell$  &   
\multicolumn{2}{c|}{$0$}         \\\hline
PRI  &  $N^2\frac{\ell-k}{k-1}$ & 
$nN/\ell$& \multicolumn{2}{c|}{$<nN(n/N)^{k-\ell}/k$ }\\\hline
\multicolumn{5}{|c|}{small item}\\\hline
U$-$R       &  \multicolumn{4}{c|}{$n/k$}  \\\hline
W$+$R       &  \multicolumn{4}{c|}{$N\ell /k$}  \\\hline
THR    &  \multicolumn{2}{c|}{$N\ell/k$} &  
\multicolumn{2}{c|}{$n/(k-\ell)$} \\\hline
PRI  &  \multicolumn{2}{c|}{$N\ell/(k-1)$} &  $n\ln N$ &
 $n/(k-\ell-1)$ \\\hline
\end{tabular}
\caption{Overview of variance with $k$ samples, in the
case of $\ell$ large items of size $N$.}\label{tab:big-small}.
\end{center}
\end{table}
\paragraph{Discussion}
With reference to Table~\ref{tab:big-small}, 
the problem with uniform sampling is that it does a terrible job
on the large weights, performing about $n/\ell$ times worse than
the other schemes. On the other hand, it gives the best performance
on the small items. However, the advantage over threshold and priority
sampling becomes insignificant when $k\gg\ell$. This illustrates
that if the number of dominant items is small compared with the number of
samples, then threshold and priority sampling do very well even on
the small items.

The problem in weighted sampling with replacement is that it does poorly
compared with threshold and uniform sampling when the number of samples 
exceed the number of dominant items. This is both large and small
items, illustrating the problem with duplicates.

Finally, comparing threshold and priority sampling, we see that
priority sampling has positive variance for $k>\ell$ whereas
threshold sampling has no variance. However, this variance of
priority sampling is very small compared to a weight of $N$, so
it is a case where priority sampling is doing very well anyway.
It is more interesting to see what happens with the small items.
The major differences are in the two boundary cases
when $k=1$ and when $k=\ell+1$. The former case has infinite variance
as discussed previously. For $k=\ell+1$, we see that priority
sampling does worse by a factor of $\ln N$. This is only by the
logarithm of the ratio of the large weight over the small weight,
and it is only for the special boundary case when $k=\ell+1$ that
we have such a big difference. It is therefore not surprising that
this kind of difference did not show up in any of our experiments. Also,
we note that in this special case, weighted sampling with replacement
is performing even much worse; namely be a factor of $N/n$. 

Thus, in our analysis, priority sampling performs very well compared
with the other schemes for sampling exactly $k$ items, and it
is only in rather singular cases that it performs a worse than threshold 
sampling.

\paragraph{Tailoring a better scheme for $k=\ell+1$ samples}
In our large-small weight example, for $k$ not too small, priority
sampling is only beaten by threshold sampling, which, however, does
not sample exactly $k$ items. In particular, priority sampling is
outperformed for $k=\ell+1$. We will now construct a sampling scheme
for this particular case which samples exactly $k$ items and gets the
same performance as threshold sampling for any $k>\ell$.  
Like threshold sampling, the tailored scheme picks all
the $k$ large items. Moreover, it picks 
$k-\ell$ items uniformly without replacement among the
small unit items. From our study of the unit case, we know
that uniform sampling gets the same variance as that
of priority sampling on the small items. Thus each item
gets the same variance with our tailored scheme as threshold
sampling, but that our tailored scheme samples exactly
$k$ items.

\section{Conjectured near-optimality of priority sampling}
\label{sec:conj-opt}
Recall from \Sref{sec:threshold} that threshold sampling
minimizes the total variance when we do independent sampling
getting an expected number of $k$ samples. We would
have liked to provide a somewhat similar result for priority
sampling among schemes sampling exactly $k$ items, but
we know that this is not the case. For unit items,
uniform sampling without replacement got an item variance of $\frac{n-k}k$ 
while priority sampling got an item variance of $\frac{n-k}{k-1}$.
Also, for our large-small item, we found a specialized scheme
outperforming priority sampling when $k>\ell$. 

We formalize our intuition as the conjecture that if priority sampling
is allowed just one extra sample, it beats any specialized
sampling scheme on any sequence of weights. More precisely,
\begin{conjecture}\label{conj} For any weight sequence $w_0,...,w_{n-1}$ and
positive integer $k\leq n$, there is no tailored scheme for picking
a sample $S\subseteq [n]$ of up to $k$ items $i$ with 
unbiased weight estimates $\hat w_i$ (that is, $\hat w_i=0$ if $i\not\in S$
and $\Ep{\hat w_i}=w_i$ for all $i\in[n]$) so that the total variance 
($\sum_{i\in[n]} \Varp{\hat w_i}$) is smaller than with a priority
sample of size $k+1$.
\end{conjecture}
The conjecture also covers tailored schemes where the same item is picked
multiple times, or where less than $k$ samples may be picked, as in
weighted sampling with replacement. If we have multiple weight estimates
for an item $i$, we add them up to a single weight estimate $\hat w_i$,
and if the sample has less than $k$ items, we add extra items $j$ with
$\hat w_j=0$. Thus the tailored scheme is transformed into one
that always picks exactly $k$ distinct items.

In fact, Conjecture\tref{conj} is equivalent to the following conjecture
relating priority sampling to threshold sampling:
\begin{conjecture}\label{conj2} For any weight sequence $w_0,...,w_{n-1}$ and
positive integer $k$, threshold sampling with an expected number of
$k$ samples gets a total variance which is no smaller than with a
priority sample of size $k+1$.
\end{conjecture}
One consequence of Conjecture\tref{conj2} is that if we only have
resources for a certain number $k$ of samples, then we are much better
off using priority sampling than using threshold sampling
for a small enough expected number of samples, e.g., $k-2\sqrt{k}$,
that the probability of getting more than $k$ samples is small.

To see that Conjectures\tref{conj} and \ref{conj2} are equivalent,
we prove
\begin{proposition} For any weight sequence $w_0,...,w_{n-1}$ and
positive integer $k$, there is no scheme for picking
a sample $S\subseteq [n]$ of $k$ items $i$ with 
unbiased weight estimates $\hat w_i$ so that the total variance 
is smaller than with threshold
sampling with an expected number of $k$ samples. In fact, given
the weight sequence, we can
construct an optimal scheme for picking $k$ items getting exactly 
the same total
variance as that of threshold sampling.
\end{proposition}
\begin{proof}
Let $\Psi$ be a scheme for picking a 
sample $S\subseteq [n]$ of $k$ items $i$ with 
unbiased weight estimates $\hat w_i^{\Psi}$. We then
consider the corresponding scheme $\cI(\Psi)$ for
independent sampling. More precisely, $I(\Psi)$ considers each
item $i$ independently, picking $i$ with the same probability
$p_i$ as does $\Psi$, and with the same probability distribution on
the weight estimate $\hat w_i^{\cI(\Psi)}$ as $\Psi$ induces
on its weight estimate $\hat w_i^{\Psi}$. Then
$\Ep{\hat w_i^{\cI(\Psi)}}=\Ep{\hat w_i^{\Psi}}=w_i$
and $\Varp{\hat w_i^{\cI(\Psi)}}=\Varp{\hat w_i^{\Psi}}$.
Moreover, by linearity of expectation, the expected
number of samples with $\cI(\Psi)$ is
\[\sum_{i\in[n]}\Prp{i\in S^{\cI(\Psi)}}=
\sum_{i\in[n]}\Prp{i\in S^{\Psi}}=k.\] 
Thus the independent sampling
scheme $\cI(\Phi)$ has unbiased estimators like $\Phi$, an expected
number of $k$ samples, and the same item variances as $\Phi$.

Now, suppose for some item $i$ that $\cI(\Phi)$ has more
than one possible non-zero weight estimate $\hat w_i^{\cI(\Psi)}$.
We then make an improved sampling scheme $\cI^*(\Phi)$ which picks
item $i$ with the same probability $p_i$ as $\Phi$ and $\cI(\Phi)$,
but which then always uses the same weight estimate
$\hat w_i^{\cI^*(\Psi)}=\Ep{\hat w_i^{\cI(\Psi)}\;|\;i\in S^{\cI(\Psi)}}$.
Then $\Ep{\hat w_i^{\cI^*(\Psi)}}=\Ep{\hat w_i^{\cI(\Psi)}}=w_i$ and
$\Varp{\hat w_i^{\cI^*(\Psi)}}\leq \Varp{\hat w_i^{\cI(\Psi)}}
=\Varp{\hat w_i^{\Psi}}$ with strict inequality if 
$\Varp{\hat w_i^{\cI(\Psi)}\;|\;i\in S^{\cI(\Psi)}}>0$. For example,
we have strict inequality if $\Psi$ is a priority sampling scheme with $k<n$.

The optimized scheme $\cI^*(\Phi)$ has the same format as the schemes
considered in \Sref{sec:threshold}, and we know that threshold
sampling minimizes the total variance among these schemes. Consequently,
with $THR$ denoting threshold sampling of an expected number of $k$ 
items, it follows that 
\[\sum_{i\in[n]} \Varp{\hat w_i^{THR}}\leq
\sum_{i\in[n]} \Varp{\hat w_i^{\cI^*(\Psi)}}
\leq \sum_{i\in[n]} \Varp{\hat w_i^{\Psi}}.\]

We will now go the other way. Our staring point is an independent
sampling scheme $\Phi$ that picks each item $i$ independently with 
probability $p_i$, and which picks an expected integer number $k$ of samples.
If item $i$ is picked, it gets weight estimate $\hat w_i^{\Phi}=w_i/p_i$;
$0$ otherwise. For example, $\Phi$ could be our threshold sampling scheme
$THR$. We will now define a corresponding sampling scheme $\cE(\Phi)$ picking
exactly $k$ samples, and getting the same variance for each item.

We are going to describe an iterative procedure defining $\cE(\Phi)$. 
Initially, set $r_i=1-p_i$ for all $i\in[n]$. We are going to define
different events, and as we do so, reduce $p_i$ and $r_i$ so as to
reflect the remaining probability that item $i$ is picked or not
picked, respectively. After each iteration, we have a remaining
total probability $P=p_0+r_0=p_1+r_1=\cdots=p_{n-1}+r_{n-1}$.
In each event we pick exactly $k$ items, and since we start
with an expected number of $k$ items, we will always have
an expected number of $k$ items in the remainder, that is
$(\sum_{i\in[n]}p_i)/P=k$.

Consider an item $i$. If $p_i=0$, item $i$ is not picked in any
remaining event. Conversely, if $r_i=0$, item $i$ is forced to
be picked in all remaining events. If $p_i>0$ and $r_i>0$, item $i$ is 
``unsettled''. If there are no unsettled events, we have a final
event, doing what has to be done: since $(\sum_{i\in[n]}p_i)/P=k$
and since each $p_i$ is either $P$ or $0$, there are exactly
$k$ items $i$ with $p_i=P$, and these are all picked.

Assume that we have some unsettled items $i$.  Let $n'$ be the number
of unsettled items. Also, let $k'$ be the number of items to be picked
among the unsettled items, that is, we subtract the forced items that
have to be picked because $r_i=0$. In our next event $A$, we want to
pick the forced items and $k'$ items uniformly from the unsettled
items.  Then item $i$ is picked with probability $k'/n'$. Hence, if
$P_A$ is the probability of the event $A$, then for each unsettled
item $i$, we will reduce $p_i$ by $P_Ak'/n'$, and $r_i$ by
$P_A(n'-k')/n'$.

We choose $P_A$ maximally, subject to the condition that no
$p_i$ or $r_i$ may turn negative. Then
\[P_A=\max\{p_in'/k',\;r_in'/(n'-k')\;|\;i\in[n],\, p_i>0,\,r_i>0\}\]
With this choice of $P_A$, the event $A$ will settle at least
one item, so it will take at most $n$ iterations to define
the sampling scheme $\cE(\Phi)$.

By definition, for each item $i$, we get the same distribution
of weight estimates with $\cE(\Phi)$ as with $\Phi$, hence also
the same variances. In particular it follows that $\cE(THR)$ has
the same total variance as does threshold sampling.

Note that when $k<n$, the total variance of $\cE(THR)$ is always smaller
than that of priority sampling since 
\begin{eqnarray*}
\sum_{i\in[n]} \Varp{\hat w_i^{\cE(THR)}}
&=&\sum_{i\in[n]} \Varp{\hat w_i^{THR}}\\
&\leq&\sum_{i\in[n]} \Varp{\hat w_i^{\cI^*(PRI)}}\\
&<&\sum_{i\in[n]} \Varp{\hat w_i^{PRI}}.
\end{eqnarray*}
For example, this was how we improved priority sampling in the special
case at the end of the previous section.
\end{proof}
As evidence for Conjecture\tref{conj2}, we note from \Sref{sec:analysis}
that it is true for the unit case. Also, it can be proved
to hold for the large-small example using a more refined analysis.
Finally, we note that the conjecture conforms nicely with the closeness of
priority sampling and threshold sampling in the experiments from 
\Sref{sec:experiments}. Also, at appears that we can prove
an asymptotic version; namely that 
$\sum_{i\in[n]} \Varp{\hat w_i^{PRI[k+1]}}\leq 
a\sum_{i\in[n]} \Varp{\hat w_i^{THR[k]}}$ where $a$ is a large enough
constant, PRI$[k+1]$ is priority  sampling of $k+1$ items, and THR$[k]$ is 
threshold sampling of an expected number of $k$ items. However, this is 
complicated, and beyond the scope of the current paper. 

Very recently, Szegedy \cite{Sze05} has settled Conjecture\tref{conj2}. By
the above equivalence, his proof also implies Conjecture\tref{conj}. Thus
priority sampling is variance optimal modulo one extra sample.

\section{Sampling from a stream}\label{sec:alg}
In this section, we will discuss how we can maintain a sample
of size $k$ for a stream of items $i=0,1,2,...$ with weights $w_i$.
\subsection{Reservoir sampling}
In so-called reservoir sampling, at any point in
time, we want to have a sample of size $k$ from the items seen so far. Thus,
if we have seen items $0,...,n-1$, we should have 
a sample $S\subseteq[n]$. The individual samples are
denoted $S[0],..,S[k-1]$. 
\paragraph{Uniform sampling with replacement} This
case was studied by Vitter \cite{Vit85}. 
Let $S^{U-R}\subseteq[n]$ be the current sample.
While $n\leq k$, we have $S[i]=i$ for
$i=0,...,n-1$.  When
item $n>k$ arrives, we pick a random number $j\in[n+1]$.
If $j<k$, we set $S^{U-R}[j]:=n$. Finally, we set
$n:=n+1$. All this takes constant time for each item.

We note that the weight estimates are only
maintained implicitly via $n$. If $j\in S^{U-R}$, 
then $\hat w_j=\frac nk w_j$ where $n$ is the current
number of items. 

\paragraph{Weighted sampling with replacement}
This case was studied by Chaudhuri et al.\cite{CMN99}.
Besides maintaining a sample $S^{W+R}\subseteq[n]$, 
we maintain the total current weight $W=\sum_{i\in[n]} w_i$.
When item $n$ arrives, for $j=0,...,k-1$, we pick
a random number $\alpha\in(0,1)$. If $\alpha\leq \frac {w_n}{W+w_n}$,
we set $S^{W+R}[j]:=n$. When done with all samples, we set
$W:=W+w_n$. Note that if we had $w_n\geq W$, we would expect
to change at least half the samples, so for exponentially
increasing weight sequences, we spend $\Theta(k)$ time on each
item. However, in \cite{CMN99}, it is falsely claimed that their
algorithm spends constant time on each item.

Using the current value of $W$, we can compute the weight estimates of
the sampled items as described in \Sref{sec:experiments}.

\paragraph{Priority sampling}
Priority sampling is trivially implemented using a standard priority
queue \cite{CLRS01}. Recall that for each item $i$, we generate a random
number $\alpha_i\in(0,1)$ and a priority $q_i=w_i/\alpha_i$.
A priority queue $Q$ maintains the $k+1$ items of
highest priority. The $k$ highest form our sample
$S$, and the smallest $q_i$ in $Q$ is our threshold $\tau$.

It is convenient to start filling our priority queue $Q$
with $k+1$ dummy items with weight and priority $0$. When
a new item arrives we simply place it in $Q$. Next
we remove the item from $Q$ with smallest priority.
With a standard comparison based priority queue, we spend
$O(\log k)$ on each item, but exploiting a floating point
representation, we can get down to $O(\log\log k)$ time
for item \cite{Tho02} (this counts the number of floating
point operations, but is independent of the precision of
floating point numbers). This is substantially better
than the $\Theta(k)$ time we spend on weighted
sampling with replacement, but a bit worse than the constant time
spent on uniform sampling without replacement. We shall
later show how to get down to constant time if we relax
the notion of reservoir sampling a bit.

\paragraph{Reservoir sampling for threshold sampling}
In \cite{DLT05}, the threshold $\tau^{THR}$ was determined
before items where considered. The threshold was adapted to
the traffic to get a desired amount of samples, yet bursts
in traffic lead to bursts in the sample. Here, as a new
contribution to threshold sampling, we present a reservoir
version of threshold sampling which at any time maintains
a sample $S^{THR}$ of expected size $k$.

As items stream by, we generate priorities as
in priority sampling. At any point, $n$ is the number of items
seen so far. We maintain a threshold $\tau^{THR}$ that would
give an expected number $k$ of items, that is, 
\begin{equation}\label{eq:thr-k}
\sum_{i\in[n]}\min\{1,w_i/\tau^{THR}\}=k
\end{equation}
Also, we maintain the corresponding threshold sample, that is,
\[S^{THR}=\{i\in[n]|q_i>\tau^{THR}\}.\]
The sample $S^{THR}$ is stored in a priority queue. When
a new item $n$ arrives it is first added to $S^{THR}$.
Next we have to increase $\tau^{THR}$ so as to satisfy 
\req{eq:thr-k} with $n'=n+1$. Finally, we remove all the items from 
$S^{THR}$ with priorities lower than $\tau^{THR}$. Thanks
to the priority queue, each such item is extracted in
$O(\log k)$ time.

We still have to tell how we compute the threshold. Together
with the sample, we store the set $L$ of all items $i$ with weight 
$w_i\geq \tau^{THR}$. Also, we store the total weight $U$ of all
smaller items. We note that the set $L$ is contained in $S^{THR}$.
Now,
\[\sum_{i\in[n]}\min\{1,w_i/\tau^{THR}\}=|L|+U/\tau^{THR}\]
The items $i$ in $L$ are stored in a priority queue ordered
not by priority $p_i$ but by weight $w_i$. When
item $n$ arrives we do as follows. If $w_i\geq \tau^{THR}$, we
add $i$ to $L$; otherwise we add its weight $w_n$ to $U$.

Next we increase $\tau^{THR}$ in an iterative process.
Let $\tau^*=U/(k-|L|)$ and let $w_j$ be the smallest
weight in $L$. If $L$ was empty, $w_j=\infty$. 
If $\tau^*<w_j$, we set $\tau^{THR}=\tau^*$, and we are done. 
Otherwise, we set $\tau^{THR}=w_j$,  remove $j$ from $L$, add
$w_j$ to $U$, and repeat.

In the above process, each item is inserted and deleted at most
once from each priority queue. Also, at any time, the expected size
of each priority queue is at most $k$, so the total expected cost per 
item is $O(\log k)$. Exploiting a floating point representation
of priorities, this can be reduced to $O(\log\log k)$ time.
Thus we get the same time complexity as for priority sampling,
but with a more complicated algorithm.

\subsection{Relaxed priority sampling}
We will now bring down the time per item to constant for
priority sampling. To do this, we relax the notion of reservoir sampling, 
and set aside space for $2k+2$ 
items. Instead of using a priority queue, we use a buffer
$B$ for up to $2k+2$ items. The buffer is guaranteed to contain
the $k+1$ items of highest priority. When it gets full, a cleanup is 
performed to reduce the occupancy down to
$k+1$.  Using a standard selection algorithm \cite{CLRS01}, we
find the $(k+1)^{\textrm{st}}$ highest priority in $B$, and all items
of lower priority are deleted, all in time linear in $k$. The cleaning
is executed once for every $k+1$ arrivals, hence
at constant cost $O(1)$ per item
processed. After cleanup, we resume filling the buffer with fresh
arrivals. 

A further modification processes every item in constant time without
having to wait for the cleanup to execute.  Two
buffers of capacity $2k+2$ are used, one buffer being used for
collection while the other is cleaned down to $m+1$ items.  Then each
item is processed in constant time, plus $O(k)$ time at the end of the
measurement period in order to find the $k+1$ items of largest
threshold from the union of the contents of the two buffers. Thus,
provided the between successive arrivals should be bounded below by
the $O(1)$ processing time per item, the processing associated with
each flow can be completed before the next flow arrives.

We note that similar ideas can be used to get constant processing
time per item for weighted sampling with replacement and threshold sampling. 

\section{Conclusions}\label{sec:conclusion}
We have introduced priority sampling as a simple scheme for
weight sensitive without replacement that is very effective for
estimating subset sums.


\begin{thebibliography}{99}


\bibitem{CLRS01}
T. H. Cormen, C. E. Leiserson, R. L. Rivest, C. Stein,
``Introduction to algorithms'', 2nd Edition, MIT Press, McGraw-Hill,
2001.



\bibitem{AB88} B.C. Arnold and N. Balakrishnan, ``Relations, Bounds
and Approximations for Order Statistics'', Lecture Notes in
Statistics, vol. 53, Springer, New York, 1988.

\bibitem{BH83} K.R.W. Brewer and M. Hanif, ``Sampling With Unequal
Probabilities'', Lecture Notes in Statistics, vol. 15, 
Springer, New York, 1983.

\bibitem{CMN99} S. Chaudhuri, R. Motwani, V.R. Narasayya: On Random
Sampling over Joins. SIGMOD Conference 1999: 263-274


\bibitem{David81} H.A. David, ``Order Statistics'', Second Edition,
  Wiley Series in Probability and Mathematical Statistics, Wiley, New
  York, 1981

 



\bibitem{DLT04} N.G. Duffield, C. Lund, M. Thorup, ``Flow Sampling
Under Hard Resource Constraints'', ACM SIGMETRICS 2004, pages 85--96.

\bibitem{DLT05}
N.G. Duffield, C.~Lund, and M.~Thorup.
\newblock Learn more, sample less: control of volume and variance in network
  measurements.
\newblock {\em IEEE Transactions on Information Theory}, 51(5):1756--1775,
  2005.


\bibitem{flow98} A. Feldmann, J. Rexford, and R. C\'{a}ceres,
``Efficient Policies for Carrying Web Traffic over Flow-Switched
Networks,'' {\em IEEE/ACM Transactions on Networking}, vol. 6, no.6,
pp. 673--685, December 1998.

\bibitem{Mut04} S. Muthukrishnan.
``Data Stream Algorithms''. Available at http://www.cs.rutgers.edu/$sim$muthu, 2004


\bibitem{PKC96} K. Park, G. Kim, and M. Crovella, "On the Relationship
  Between File Sizes, Transport Protocols, and Self-Similar Network
  Traffic". In Proc. 4th International Conference on
  Network Protocols (ICNP), pp. 171--180, 1996).


\bibitem{Slammer} 
  http://securityresponse.symantec.com/avcenter/venc/data/w32.sqlexp.worm.html


\bibitem{Sze05}
Mario Szegedy.
\newblock Near optimality of the priority sampling procedure.
\newblock Technical Report TR05-001, Electronic Colloquium on Computational
  Complexity, 2005.



\bibitem{Tho02} M. Thorup, "Equivalence between Priority Queues and
  Sorting", Proc. 43rd IEEE Symposium on Foundations of
  Computer Science (FOCS), pp. 125--134, 2002.


\bibitem{Vit85} J.S. Vitter: Random Sampling with a Reservoir. 
{\em ACM Trans. Math. Softw.} 11(1): 37-57 (1985)

\bibitem{WE80} C.K. Wong, M.C. Easton: An Efficient Method for 
  Weighted Sampling without Replacement. {\em SIAM J. Computing} 9(1):
  111--113 (1980)




\end{thebibliography}
\end{document}